\documentclass[pra,reprint,amssymb,superscriptaddress]{revtex4-1}

\usepackage{natbib}
\usepackage{gensymb}
\usepackage{amsthm}
\usepackage{amsmath}
\usepackage{color}
\usepackage{bm}
\usepackage{amsmath,amssymb}
\usepackage{amsfonts}
\usepackage{dsfont}
\usepackage{graphicx}
\usepackage{float}
\usepackage{subfigure}
\usepackage{verbatim}
\usepackage{amsfonts}
\usepackage{bbold}
\usepackage{dcolumn}
\usepackage{bm}
\usepackage[dvipsnames]{xcolor}
\usepackage{listings}
\definecolor{blueprl}{RGB}{46,48,146}
\usepackage[pdftex,colorlinks=true,urlcolor=blueprl,citecolor=blueprl,linkcolor=blueprl]{hyperref}
\usepackage{mathrsfs}
\usepackage{filecontents}
\usepackage{mathtools}
\usepackage{times}
\usepackage{color}
\usepackage[dvipsnames]{xcolor}

\usepackage{braket}

\usepackage{soul}

\date{\today}

\begin{document}
\title{Noise Transfer Approach to GKP Quantum Circuits}
\author{Timothy C. Ralph}
\email{ralph@physics.uq.edu.au}
\affiliation{Centre for Quantum Computation and Communication Technology, School of Mathematics
and Physics, University of Queensland, Brisbane, Queensland 4072, Australia}

\author{Matthew S. Winnel}
\email{mattwinnel@gmail.com}
\affiliation{Centre for Quantum Computation and Communication Technology, School of Mathematics
and Physics, University of Queensland, Brisbane, Queensland 4072, Australia}

\author{S. Nibedita Swain}
\affiliation{Centre for Quantum Computation and Communication Technology, School of Mathematics
and Physics, University of Queensland, Brisbane, Queensland 4072, Australia}

\affiliation{School of Mathematical and Physical Sciences,
University of Technology Sydney, Ultimo, NSW 2007, Australia}

\affiliation{Sydney Quantum Academy, Sydney, NSW 2000, Australia}

\author{Ryan J. Marshman}
\affiliation{Centre for Quantum Computation and Communication Technology, School of Mathematics
and Physics, University of Queensland, Brisbane, Queensland 4072, Australia}

\begin{abstract}
The choice between the Schr\"{o}dinger and Heisenberg pictures can significantly impact the computational resources needed to solve a problem, even though they are equivalent formulations of quantum mechanics.
    Here we present a method for analysing Bosonic quantum circuits based on the Heisenberg picture that allows, under certain conditions, a useful factoring of the evolution into signal and noise contributions, in a similar way as can be done with classical communication systems. We provide examples which suggest this approach may be particular useful in analysing quantum computing systems based on the Gottesman-Kitaev-Preskill (GKP) qubits.
\end{abstract}

\maketitle

\section{Introduction}\label{sec:intro}

A useful technique for analysing quantum optical experiments based on Gaussian operations and measurements is the quantum noise transfer approach \cite{BAC19} which is characterised by the displacement vector and the covariance matrix \cite{WEE12}.  Working in the Heisenberg picture, the signals (the displacements) and the quantum noise (the covariances) can be tracked separately, in much the same way that signals and noise are tracked, for example,  in classical communication systems.  A key step in such an analysis is the definition of quadrature fluctuation operators via:
\begin{equation}
\delta \hat q = \hat q -  q ,\;\;\;
\delta \hat p = \hat p -  p ,
\label{fopg}
\end{equation}
where $q=\langle \hat q \rangle$ and $p = \langle \hat p \rangle$ are real numbers representing the average values of the position and momentum quadrature displacements respectively.  Also, $\hat q$ and $\hat p$ are the position and momentum operators for the mode in question, respectively.  We use the convention that the annihilation operator for the mode is given by $\hat a = 1/2(\hat q + i \hat p)$ which can then be expanded as:
\begin{equation}
\hat a = 1/2(q + \delta \hat q + i(p + \delta \hat p)),
\end{equation}
and the classical signals (displacements) and the quantum noise operators can be independently tracked in the Heisenberg picture through various interactions and measurements in an intuitive way. In an experimental setting the noise properties of the input states can be measured directly by subtracting off the known signal and evaluating the variance of the quantum fluctuations. For example the position quadrature variance is:
\begin{equation}
V_q \equiv \langle \delta \hat q^2 \rangle = \langle (\hat q -  q)^2 \rangle = \langle \hat q^2 \rangle - q^2
\end{equation}

The aim of this paper is to develop an analogous approach that can describe the evolution of non-Gaussian states, such as cat states \cite{RAL03} or Gottesman-Kitaev-Preskill (GKP) states \cite{GOT01}, in a similar way.  Such states are formed from a superposition of differently displaced states and hence the signal is multi-valued.  

In the next section we will introduce our new decomposition into signal and fluctuation operators and show how noise properties can be analysed independently of the signal value provided certain conditions are met.  We illustrate this with cat state and GKP state examples. In Section 3 we consider the Heisenberg Picture evolution of our operators with loss and feedforward as examples. In Section 4 we apply our formalism to
GKP states and 
analyse a teleportation channel which includes the GKP error correction protocol. The goal is to evaluate the noise transfer properties of the circuits based on the first and second moments of the resource states (and the properties of the feedforward measurements) -- thus providing an 
alternative method for analysing quantum computing circuits based on GKP qubits. Current approaches to GKP circuit evaluation are mostly based on Schrödinger picture analysis. Noise can be added to ideal GKP states \cite{MEN14} and the logical states tracked through a modular sub-system decomposition \cite{PBM20} which can be generalised to finite energy GKP states \cite{TZI20,BOU21}. Exact numerical models incorporating noisy elements have also been explored \cite{HIL22}. In contrast the method presented here differs by describing a way to empirically quantify the noise and signals independently and then track them through circuits via their Heisenberg evolution, offering potential advantages in the intuitive nature of the approach and the tractability of the calculations in the presence of multiple noise sources. 
We discuss our results and conclude in Section 5.

\section{Signal and fluctuation Operators}

We start by defining the operators $\delta \hat q$ and $\delta \hat p$ in an analogous way to Eq.\ref{fopg} via:
\begin{equation}
\delta \hat q = \hat q - \hat q_c, \;\;\;
\delta \hat p= \hat p - \hat p_c.
\end{equation}
Because the displacements are multi-valued they must be represented by operators, specifically $\hat q_c$ and $\hat p_c$.  
We characterise $\hat q_c$ and $\hat p_c$ by their moments over restricted domains around their expected displacement values in the position and momentum quadratures respectively. Specifically,  given $q_n = \langle \hat q \rangle_n$ is the expectation value for $q$ values falling in the $n$th position domain, we require that $\langle \hat q_c \rangle_n = q_n$.  We further require that $\langle \hat q_c^2 \rangle_n = q_n^2$, meaning that $ \hat q_c$ has zero variance in each domain.  This implies $\langle \hat q \hat q_c \rangle_n = q_n^2$.  Similarly we require $\langle \hat p_c \rangle_n = p_n$ where $p_n = \langle \hat p \rangle_n$ is the expectation value for $p$ values falling in the $n$th momentum domain, $\langle \hat p_c^2 \rangle_n = p_n^2$ and hence $\langle \hat p \hat p_c \rangle_n = p_n^2$.  Assuming we have $N_q$ position domains and $N_p$ momentum domains, the full expectation values are given by
\begin{eqnarray}
\langle \hat q_c \rangle &=&  \sum_n^{N_q} q_n P_{nq}, \\ \nonumber
\langle \hat p_c \rangle &=&  \sum_n^{N_p} p_n P_{np}
\end{eqnarray}
with $P_{nq}$ the probability of finding a $q$ value in the $n$th domain and $P_{np}$ the probability of finding a $p$ value in the $n$th domain.

Our mode operators are now of the form:
\begin{equation}
\hat a = 1/2(\hat q_c + \delta \hat q + i(\hat p_c + \delta \hat p)),
\label{mode}
\end{equation}
The noise properties of the input states can be measured directly by subtracting off the closest expected displacement, evaluating the variance of the quantum fluctuations around this expected value,  and taking the weighted average over all expected values. These quantities can then be related to $\delta \hat q$ and $\delta \hat p$ in the following way.  
We can write the average position quadrature variance described above as:
\begin{eqnarray}
V_q &\equiv & \sum_n^{N_q} (\langle \hat q^2 \rangle - q_n^2) P_{nq} \\ \nonumber
&=& \langle \hat q^2 \rangle - \sum_n^{N_q} q_n^2 P_{nq}  \\ \nonumber
&=& \langle \hat q^2 \rangle - \sum_n^{N_q}(2 \langle \hat q \hat q_c \rangle_n - \langle \hat q_c^2 \rangle_n)P_{nq}  \\ \nonumber
&=&  \langle (\hat q -  \hat q_c)^2 \rangle  \\ \nonumber
&=&  \langle \delta \hat q^2 \rangle.
\end{eqnarray}
This can be evaluated via
\begin{eqnarray}
V_q &=& \int dq \; q^2 |\Psi(q)|^2 - \sum_n^N {{(\int_n dq \; q |\Psi(q)|^2)^2}\over{(\int_n dq |\Psi(q)|^2)}}.
\label{vq}
\end{eqnarray}
where we have used:
\begin{eqnarray}
q_n &=& {{\int_n dq \; q |\Psi(q)|^2}\over{\int_n dq |\Psi(q)|^2}}; \\ \nonumber
P_{nq} &=& \int_n dq |\Psi(q)|^2,
\end{eqnarray} 
and the integral $\int_n dq$ is taken over the corresponding $n$th domain whilst $\Psi(q) = \langle q| \psi \rangle$ is the position wavefunction of the state $| \psi \rangle$. Similarly,   the momentum quadrature variance can be written:
\begin{eqnarray}
V_p &\equiv & \langle \delta \hat p^2 \rangle = \langle (\hat p -  \hat p_c)^2 \rangle  \\ \nonumber
&=& \langle \hat p^2 \rangle - \langle \hat p_c^2 \rangle \\ \nonumber
&=& \langle \hat p^2 \rangle - \sum_n^{N_p} p_n^2 P_{np}.
\end{eqnarray}
This can be evaluated via 
\begin{eqnarray}
V_p &=& \int dq \; p^2 |\Psi(p)|^2 - \sum_n {{(\int_n dp \; p |\Psi(p)|^2)^2}\over{(\int_n dp |\Psi(p)|^2)}},
\label{vp}
\end{eqnarray}
where the integral $\int_n dp$ is taken over the corresponding $n$th domain whilst $\Psi(p) = \langle p| \psi \rangle$ is the momentum wavefunction of the state $| \psi \rangle$. 

If the signal peaks are well localized in their expected domains then Eqs \ref{vq} and \ref{vp} will well represent the average spread of the signal peaks.  Assuming Gaussian statistics we can then estimate the probability that a quadrature measurement will find a result in their expected domain via
\begin{equation}
P_j = Erf[{{D}\over{2 \sqrt{2 V_j}}}],
\label{error}
\end{equation}
where $D$ is the width of the domains and $j = q, p$.
 
The interpretation becomes more subtle if the variances approach the width of the domains. If this happens the probability distribution within the domain will be ``clipped", and will be changed relative to its ``unclipped" value.  We will observe this effect in our examples.

\begin{figure}
\centering
\includegraphics[width=0.7\linewidth]{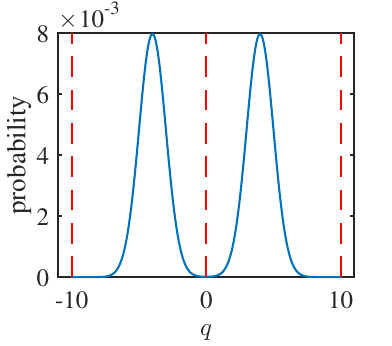}
 \caption{Example $q$ quadrature probability distribution for the cat state in Eq.\ref{cat} with $ \alpha = 2 $.}\label{fig:catm}
\end{figure}

\begin{figure}
\centering
\includegraphics[width=0.7\linewidth]{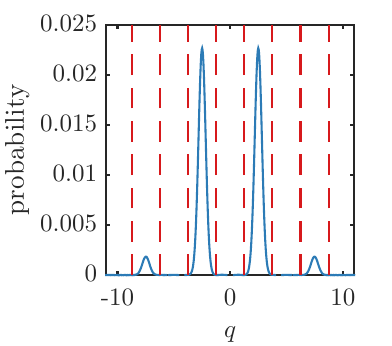}
 \caption{Example $q$ quadrature probability distribution for the GKP state in Eq.\ref{eq:GKP} with $ \mu = 1 $ and $\Delta^2 = 0.1$.}\label{fig:gkp1}
\end{figure}

\begin{figure}
\centering
\includegraphics[width=0.7\linewidth]{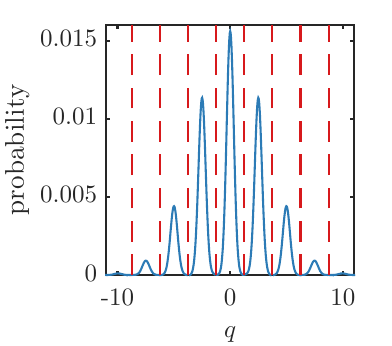}
 \caption{Example $q$ quadrature probability distribution for the GKP state in Eq.\ref{eq:GKP} with $ \mu = 1 $ and $\Delta^2 = 0.1$, but rotated through a quadrature angle of $\pi/2$. This is equal to the ``-" GKP state or equivalently the $p$ quadrature probability distribution of the ``1" state.}\label{fig:gkp+}
\end{figure}

The next two subsections will examine the examples of cat states and GKP states. `Cat state' usually refers to a superposition of differently displaced Gaussian states. The case we consider here is a symmetric superposition of 2 coherent states around the origin.  The GKP state is a particular superposition of multiple displaced Gaussian states that has the feature that if qubit values ``0" or ``1" are encoded by the displacement values in the q-quadrature then the p-quadrature encodes the ``+" and ``-" diagonal states respectively. This nice feature means that a logical Hadamard gate can be enacted on the encoded states by a simple phase rotation. The specific GKP state construction we use here is a multiple, weighted superposition of squeezed states, displaced (and squeezed) in the q-direction. Example $q$ quadrature probability distributions for cat and GKP states are shown in Figs \ref{fig:catm}, \ref{fig:gkp1} and \ref{fig:gkp+}.

\subsection{Cat States}

\begin{figure}
\centering
\includegraphics[width=0.9\linewidth]{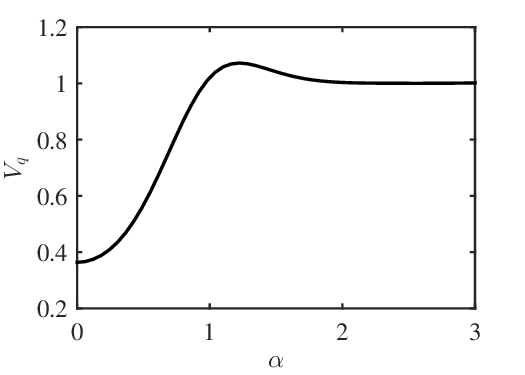}
 \caption{Average position quadrature variance $ V_q $ as a function of the parameter $ \alpha $ for the cat state defined in Eq.~\ref{cat}. Notably, $ V_q < 1 $ for small values of $ \alpha $, which can be attributed to clipping effects.}\label{fig:cat}
\end{figure}

Let us us begin with perhaps the simplest non-trivial example; a superposition of two displacements of the vacuum. In particular let us consider the cat state given by:
\begin{equation}
|\psi_c \rangle = \aleph (| \alpha \rangle + |-\alpha \rangle),
\label{cat}
\end{equation}
where $\aleph = (2 + 2 e^{-2 \alpha^2})^{-1/2}$ is a normalization constant and $|\pm \alpha \rangle$ are coherent states with real displacements $\pm \alpha$.  Consider the position quadrature variance.  The wave function is 
\begin{equation}
\Psi_q = \aleph (2 \pi)^{-1/4}(e^{-(q-2 \alpha)^2/4} + e^{-(q+2 \alpha)^2/4}).
\end{equation}
We can take the domain for $n=1$ as $-\infty \to 0$ and for $n=2$ is $0 \to \infty$, leading to the 2 corresponding values for $q_n$ as 
\begin{eqnarray}
q_1 &=& {{\int_{- \infty}^0 dq \; q |\Psi(q)|^2}\over{\int_{- \infty}^0 dq \; |\Psi(q)|^2}} \approx -2 \alpha \\ \nonumber
q_2 &=& {{\int_0^{ \infty} dq \; q |\Psi(q)|^2}\over{ \int_0^{ \infty} dq \;|\Psi(q)|^2}} \approx 2 \alpha,
\end{eqnarray}
where the approximate equalities are satisfied for $|\alpha| > 1$.  The probabilities are given by
\begin{eqnarray}
P_{1q} &=&  \int_{- \infty}^0 dq \; |\Psi(q)|^2 = 1/2,  \\ \nonumber
P_{2q} &=&  \int_0^{ \infty} dq \;|\Psi(q)|^2 = 1/2.
\end{eqnarray}
Substituting these results into our expressions for $V_q$ we find that provided $|\alpha| > 1$ then $V_q \approx 1$.  This corresponds to our intuitive picture of the cat state as comprising two peaks at $\pm \alpha$ with widths of one unit of quantum noise.  However, if $|\alpha|$ falls significantly below $1$ the distributions in the two domains are only weakly peaked (if at all) and highly skewed and clipped, so the decomposition into a discrete superposition is no longer useful. This is illustrated in Fig.\ref{fig:cat}.

Even though the superposition of displacements is only explicit in the position quadrature, interference effects also lead to distinct peaks in the momentum quadrature, around which we can define domains. The momentum wave function is:
\begin{equation}
\Psi_p = (1+ e^{-8 \alpha^2})^{-1/2}( \pi/2)^{-1/4} e^{-p^2/4} \cos{2 p \alpha},
\end{equation} 
so we can define the expected value in the $n$th domain as
\begin{eqnarray}
p_n &=& {{\int_{d_{n,-}}^{d_{n,+}} dp \; p |\Psi(p)|^2}\over{\int_{d_{n,-}}^{d_{n,+}} dp \;|\Psi(p)|^2}} \approx {{n \pi}\over{2 \alpha}},
\end{eqnarray}
with corresponding probability
\begin{eqnarray}
P_{np} &=&  \int_{d_{n,-}}^{d_{n,+}} dp \;|\Psi(p)|^2
\end{eqnarray}
where $d_{n,\pm}=(n \pm 1/2) \pi/(2 \alpha)$.
Substituting these results into our expressions for $V_p$ we find the variances scale inversely with $|\alpha|$. This is unsurprising given the size of the domains are inversely proportional to $|\alpha|$.  In particular, the ratio of domain size (equivalently peak separation) to standard deviation, ${{n \pi}\over{2 \alpha}}/\sqrt{V_p}$, is roughly constant ($\approx 5.7$) for $|\alpha| > 1$.

\subsection{GKP States}

We now consider the case of GKP states.  A physical GKP ``0" state has $q$ quadrature outcome probabilities with peaks around the values $2 n \sqrt{2 \pi}$,  where $n$ is any integer and the peaks are weighted by a suitable envelope function.  The corresponding GKP ``1" state has $q$ quadrature outcome probabilities peaked around the values $2 (n+1/2) \sqrt{2 \pi}$, similarly weighted.  The $p$ quadrature outcome probabilities for these computational states have peaks at both the ``0" and ``1" positions., i.e. at values $n \sqrt{2 \pi}$.  This is consistent with the observation that a logical Hadamard gate should be implemented by a ${{\pi}\over{2}}$ delay, which takes $\hat q \to \hat p$ and $\hat p \to -\hat q$. Thus, the logical values of the $p$ quadrature for the ``0" and ``1" computational basis states correspond to ``+" and ``-" dual basis states, respectively.

We will consider a physical example of such a GKP state to see what the noise properties are for such a state.  One such example is the squeezed state superpositions \cite{TZI20}:
\begin{eqnarray}\label{GKP}
|\psi_\mu \rangle = N \sum_{n=-\infty}^{\infty} && e^{-{{\pi}\over{2}}\Delta^2(2n + \mu)^2 } \nonumber \\
&& \hat D\left(\sqrt{{{\pi}\over{2}}}(2n + \mu)\sqrt{1 - \Delta^4}\right)|\Delta \rangle, \nonumber \label{eq:GKP} \\
\end{eqnarray}
where $N$ is a normalization factor and $|\Delta \rangle$ is a squeezed state with squeezed variance $\Delta^2 < 1$.  The value of $\mu$ determines the logical state, i.e. $|\psi_0 \rangle$ is the logical ``0" state and  $|\psi_1 \rangle$ is the logical ``1" state.  
We can evaluate the noise properties of these initial states using Eq.~\ref{vq} and by defining the domains via: $ (n+1/2) \sqrt{2 \pi} < q < (n+3/2) \sqrt{2 \pi}$ and similarly for $p$.

\begin{figure}
\centering
\includegraphics[width=0.9\linewidth]{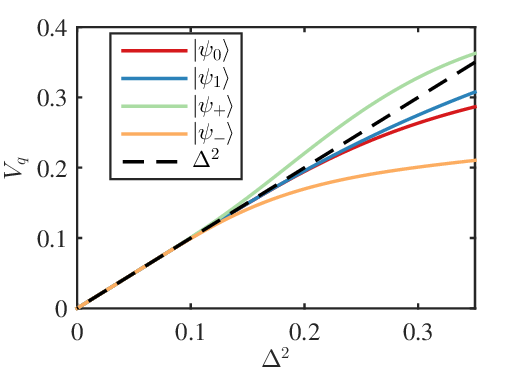}
 \caption{Average position quadrature variance $V_q$ as a function of the squeezing parameter $\Delta^2$ for GKP logical states. The computational-basis states are defined in Eq.~\ref{eq:GKP}, and the dual-basis states are simply rotated versions of the computational-basis states. The dashed line represents $\Delta^2$. $V_q$ matches $\Delta^2$ for small values of $\Delta$ but deviates in a state dependent way for larger values. Plotting $V_p$ follows a similar approach, as the $p$ quadrature is simply a rotation, with the computational and dual-basis states switching roles.}\label{fig:GKP}
\end{figure}

In Fig.\ref{fig:GKP} we plot $V_q$ as a function of $\Delta^2$ for the computational states and the diagonal states. We find that provided $\Delta^2 \lesssim {{1}\over{10}}$ then to an excellent approximation
\begin{equation}\label{GN}
V_q = V_p = \Delta^2.
\end{equation}
This is true for all logical states. This indicates that, provided the squeezing is sufficiently strong, these states behave as hoped with the superposed ``spikes" giving the logical value, modulated by Gaussian noise with a variance equal to the squeezing, in both quadratures.  Alternatively, given that this is an expected result for GKP states \cite{GOT01}, one can take this as indicating that our Heisenberg approach can successfully factor the operators into signal and noise parts in a consistent way for such states.  However, at lower levels of squeezing (i.e. larger $\Delta$) state-dependent effects are seen. For the computational states this mostly occurs due to ``clipping" of the distribution by the domain boundary such that the calculated $V_q$ no longer aligns well with the actual variance around the spikes. On the other hand the deviation of the diagonal states from the expected behaviour is predominantly due to our approximate GKP states no longer exhibiting the expected noise symmetry between the spikes in different quadratures.

\section{Operator Evolution and Feedforward}

%

Recall that our mode operators can be written in the form of Eq.~\ref{mode}.
Using operators of this kind to represent the various input modes that interact in an optical circuit we can proceed to evolve them through beam-splitters, squeezers, and other quadratic unitaries (linear in the mode operators) in the usual way, whilst keeping track of their noise and signal properties.  For example, suppose our initial mode, Eq.~\ref{mode}, passes through loss, such that only the fraction $\eta$ is transmitted. Our output mode is
\begin{equation}
\hat a_l = 1/2( \sqrt{\eta}(\hat q_c + \delta \hat q + i(\hat p_c + \delta \hat p))+\sqrt{1-\eta}(\hat q_v +i \hat p_v)),
\label{mode l}
\end{equation} 
where $\hat q_v$ and $\hat p_v$ are the position and momentum quadrature operators of the vacuum mode introduced by the loss, respectively.  By inspection we see that the expected signal values (and hence corresponding domain boundaries) have been scaled by the factor $\sqrt{\eta}$.  On the other hand the variances around these expected values are now 
\begin{eqnarray}
V_{ql} &=& \eta \langle \delta \hat q^2 \rangle +(1-\eta) \langle \delta \hat q_v^2 \rangle  \\ \nonumber
&=& \eta V_q + (1- \eta),
\end{eqnarray}
and
\begin{eqnarray}
V_{pl} &=& \eta \langle \delta \hat p^2 \rangle +(1-\eta) \langle \delta \hat p_v^2 \rangle  \\ \nonumber
&=& \eta V_p + (1- \eta),
\end{eqnarray}
where we have used that the vacuum noise is independent and has unit variance. Considering our cat state example from the previous section we see that there is no effect on the position quadrature variance as we still have $V_{ql} =1$.  The detrimental effect of loss on position only arises from the scaling down of the expected values closer to the domain boundary at $0$. On the other hand, given that for $|\alpha|>1$, $V_p << 1$,  the effect on the momentum quadrature is a significant broadening of the peaks for relatively small amounts of loss. This combined with the reduction in the peak separation rapidly washes out the interference fringes entirely, especially for $|\alpha|>>1$. This illustrates the well-known fragility of cat states to loss.

As well as optical unitary evolution, many quantum circuits involve quadrature measurements followed by feedforward to other modes in the circuit. In particular this can occur in teleportation scenarios arising in error correction \cite{GLA06} or cluster state \cite{MEN06} protocols, or both \cite{MEN14}.  Feed forward of quadrature measurements can be represented in the usual way \cite{BAC19} by feeding forward some function of the measurement operator.
Thus a typical output mode $\hat a_o$ after a teleportation type circuit might be written formally as:
\begin{equation}
\hat a_o = 1/2(\hat q_{co} + \delta \hat q_{o} + i(\hat p_{co} + \delta \hat p_o))+ G_1 (\hat q_{1}) +i G_2 (\hat p_{2}),
\end{equation}
where feedforward from an earlier position measurement ($\hat q_{1} $) and an earlier momentum measurement ($\hat p_{2}$),  have been incorporated.  In standard teleportation, the ``$G$" functions would be linear functions of the measurement operators, i.e.  $G_1 (\hat q_{1}) = g_1 \hat q_{1}$ and $G_2 (\hat p_{2}) = g_2 \hat p_{2}$. However, in a situation where we can usefully write our measurement operators as $\hat q_{1} = \hat q_{c1} + \delta \hat q_{1}$ and $\hat p_{2}  = \hat p_{c2} + \delta \hat p_{2}$, we can perform error correction by shifting to the nearest ``ideal" result in each domain,  represented by $\hat q_{c1}$ or $\hat p_{c2}$.  Hence we now have the highly non-linear feed forward functions $G_1 (\hat q_{1}) = g_1 \hat q_{c1}$ and $G_2 (\hat p_{2}) = g_2 \hat p_{c2}$ and our output mode is now written formally as:
\begin{equation}
\hat a_o = 1/2(\hat q_{co} + \delta \hat q_{o} + i(\hat p_{co} + \delta \hat p_o))+ g_1 \hat q_{c1} +i g_2 \hat p_{c2}.
\end{equation}
To illustrate feed forward with error correction we will consider a simple GKP circuit. 
\begin{figure}[htb]
\begin{center}
\includegraphics*[width=7cm]{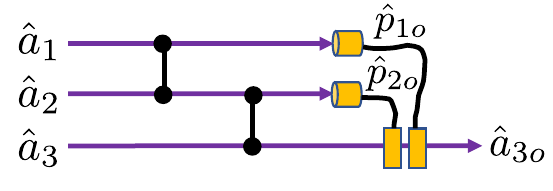}
\caption{Simple teleportation circuit with CZ gates to interact the modes and feedforward of momentum measurements of mode 1 as imaginary displacements of mode 3 and momentum measurement of mode 2 as real displacements of mode 3. The measurement of mode 1 is represented by the operator $\hat p_{1o}$, but if error correction is being implemented it is $\hat p_{c1o}$ which is fed forward. Similarly,  the measurement of mode 2 is represented by the operator $\hat p_{2o}$, but if error correction is being implemented it is $\hat p_{c2o}$ which is fed forward. 
}
\label{CZ}
\end{center} 
\end{figure}

\section{GKP Error Correction}

Another useful feature of GKP states is that small displacements of the state, as will naturally arise in the presence of loss or thermal noise, can be corrected by a straightforward circuit. This process is often referred to as GKP error correction. In the following we will analyse such a GKP error correction circuit.

The circuit we will consider is shown in Fig.\ref{CZ}. It is an example of a continuous variable CZ teleportation circuit with error correction, where we assume both resource states are approximate GKP states and we neglect loss.  We will label the approximate GKP input and two resource states with the subscripts ``$1, 2, 3$" respectively.  The resource states in modes ``2" and ``3" are prepared in the logical ``+" state, whilst the input in mode ``1" is in an arbitrary GKP logical state.

\subsection{Ideal Case}

The action of a continuous variable CZ gate is to displace the value of the $\hat p$ quadrature of one mode by the value of the $\hat q$ quadrature of the other mode, whilst leaving the $\hat q$ quadratures unchanged~\cite{MEN06}. Hence, for the circuit of Fig.\ref{CZ} the $\hat p$ quadratures will evolve such that $\hat p_1 \to \hat p_1 + \hat q_2$, $\hat p_2 \to \hat p_2 + \hat q_1 + \hat q_3$ and $\hat p_3 \to \hat p_3 + \hat q_2$. Using these expressions, and the error correction feed forward relations previously introduced, we can write the expression for the output mode as:
\begin{eqnarray}\label{3o}
\hat a_{3o} &=& {{1}\over{2}}(\hat q_{c3} + \delta \hat q_3 + i(\hat p_{c3} + \delta \hat p_3))+  \nonumber \\
&& {{i}\over{2}}(\hat q_{c2} + \delta \hat q_2 ) - {{i}\over{2}} p_{c1o} - {{1}\over{2}} p_{c2o},
\end{eqnarray}
where we have used that 
\begin{eqnarray}\label{MO}
\hat p_{1o} &=&  \hat p_{c1} +  \delta \hat p_{1} + \hat q_{c2} + \delta \hat q_{2}, \nonumber \\
\hat p_{2o} &=& \hat p_{c2} + \delta \hat p_{2} + \hat q_{c1} + \delta  \hat q_{1}  +  \hat q_{c3} + \delta \hat q_{3},
\end{eqnarray}
and therefore
\begin{eqnarray}
 G(\hat p_{1o}) &=& \hat p_{c1o}, \nonumber \\
 G(\hat p_{2o}) &=& \hat p_{c2o}.
\end{eqnarray}
If we assume that the noise terms are sufficiently small that it is unlikely that the noise causes measured values to leave their nominal domains, then we can approximate the feedforward terms as
\begin{eqnarray}\label{FF}
\hat p_{c1o} &\approx &  \hat p_{c1} + \hat q_{c2}, \nonumber \\
\hat p_{c2o} &\approx & \hat p_{c2} + \hat q_{c1} +  \hat q_{c3}.
\end{eqnarray}
and hence
\begin{eqnarray}\label{3oc}
\hat a_{3o} & = & {{1}\over{2}}(\delta \hat q_3 + i \delta \hat p_3 + i \delta \hat q_2 + \hat p_{c2}) - {{1}\over{2}}( \hat q_{c1} + i \hat p_{c1}).~~~~
\end{eqnarray}
The logical state of the output matches that of the input as seen from the output signal terms $ \hat q_{c1} + i \hat p_{c1}$. Notice that because the resource states are prepared in the ``+" state, the term in the output signal $ \hat p_{c2}$ has a known logical ``0" value and so its addition to the output does not change the overall logical value. Three noise terms are also added to the output.

The circuit can be iterated by taking $\hat a_{3o}$ as the mode 1 of the next circuit and introducing new resource states at mode 2 and mode 3.  The output mode and variances separate conveniently into deterministic displacement terms and noise terms (the ``$\delta$" terms).  The error correction feature of the circuit can be seen from the fact that none of the noise terms in the output depend on the input. That is, even though there is still noise on the output, the noise does not ``build up" as we iterate but rather is ``refreshed" from new resource states used at each round.

\subsection{Loss Tolerance}

\begin{figure}[htb]
\begin{center}
\includegraphics*[width=8cm]{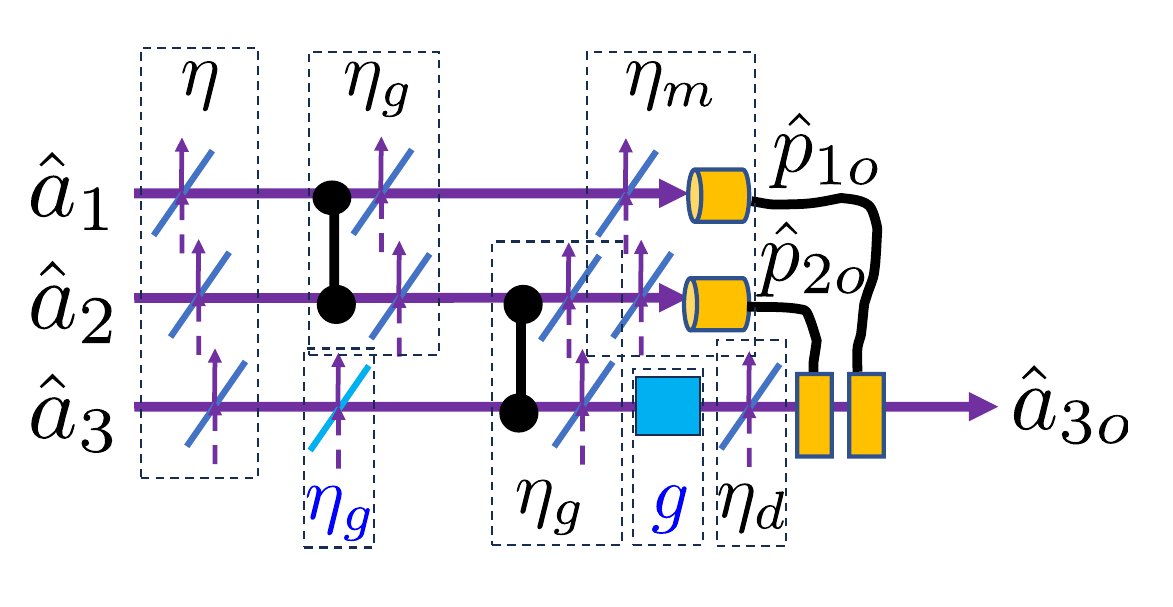}
\caption{The simple teleportation error correction circuit of Fig.3 but now with loss errors included on all components. The loss is modelled with beamsplitters where the transmission of the beamsplitters represents the efficiency of the corresponding components. Additional components (loss and linear amplification of mode 3) are indicated in blue. These components along with tailored feedforward gains allow the circuit to still implement error correction. The measurement of mode 1 is represented by the operator $\hat p_{1o}$, but if error correction is being implemented it is $\hat p_{c1o}$ which is fed forward. Similarly,  the measurement of mode 2 is represented by the operator $\hat p_{2o}$, but if error correction is being implemented it is $\hat p_{c2o}$ which is fed forward.
}
\label{CZ2}
\end{center} 
\end{figure}

We now consider the situation depicted in Fig.\ref{CZ2} in which loss effects the input state and the resource states and all the elements in the error correction circuit. We use our Heisenberg approach to demonstrate that the error correction properties of the circuit are loss tolerant, provided we add an additional loss element and a linear amplifier in appropriate positions on the third rail.

As shown in Fig.\ref{CZ2} loss is added to the input and resource states. For simplicity we use the same beamsplitter transmission value, $\eta$, for all the modes. We model loss in the CZ gates by placing beamsplitters with transmission $\eta_{\text{g}}$ after each gate. Similarly we model detection inefficiency by placing beamsplitters with transmission $\eta_{\text{m}}$ before the detectors and loss in the displacement operations by placing a beamsplitter of transmission $\eta_{\text{d}}$ before the displacements. In order to balance the loss, the experimenter should purposely add an additional beamsplitter with transmission $\eta_{\text{g}}$ to the third mode before the CZ gate. Finally, to ensure the output is balanced and centred in the domains a linear amplifier of gain $g$ is applied to the third rail after the CZ gate.

To see how this works we first write the evolved operators describing the outcomes at the first and second momentum measurements as:
\begin{align}
 \hat{p_{1o}} & = \sqrt{\eta} \sqrt{\eta_{\text{g}}} \sqrt{\eta_{\text{m}}} \Big(\hat{p}_{c1}  + \delta \hat{p}_{1} +  \hat{q}_{c2} +  \delta \hat{q}_{2} \Big) \nonumber \\
 & + \sqrt{\eta_{\text{m}}} \sqrt{1-\eta_{\text{g}} \eta}\;(\delta \hat{q}_{v2} + \delta \hat{p}_{v1} ) +  \sqrt{1-\eta_{\text{m}}}\delta \hat{p}_{m1} \\
\hat{p_{2o}} & = \eta_{\text{g}} \sqrt{\eta}  \sqrt{\eta_{\text{m}}} \Big(\hat{p}_{c2}  + \delta \hat{p}_{2} +  \hat{q}_{c1} +  \delta \hat{q}_{1} +  \hat{q}_{c3} +  \delta \hat{q}_{3} \Big) \nonumber \\ 
  & + \sqrt{\eta_{\text{m}}} \sqrt{1-\eta_{\text{g}}^2 \eta}\;(\delta \hat{q}_{v1} + \delta \hat{q}_{v3} + \delta \hat{p}_{v2} ) \nonumber \\ 
  & +  \sqrt{1-\eta_{\text{m}}}\delta \hat{p}_{m2}. 
\end{align}
Here the $\delta \hat{p}_{vi}$ and $\delta \hat{q}_{vi}$ are vacuum operators for the momenta and position respectively arising from the input mode and gate loss. The $\delta \hat{p}_{mi}$ are vacuum operators for the momenta arising from detector inefficiency.

Consider the first momentum measurement. Using the error correction strategy and a feedforward gain of $g_1 = -{{1}\over{\sqrt{\eta \eta_{\text{g}} \eta_{\text{m}}}}}$ we find the feedforward operators are approximately:
\begin{align}
     \hat{p}_{c1o} & \approx  \hat{p}_{c1} +   \hat{q}_{c2}  
\end{align}
This assumes that the variance of the feedforward operators:
\begin{align}
    V_1 = 2 \Delta^2 + 2({{1}\over{\eta \eta_{\text{g}}}}-1) + {{1}\over{\eta \eta_{\text{g}}}}({{1}\over{\eta_{\text{m}}}}-1),
\end{align}
is sufficiently small.

Now we consider the second momentum measurement. Using the error correction strategy and a feedforward gain of $g_2 = -{{1}\over{\sqrt{\eta \eta_{\text{g}}^2 \eta_{\text{m}}}}}$ we find the feedforward operators are approximately:
\begin{align}
      \hat{p}_{c2o} & \approx  \Big(\hat{p}_{c2}  + \hat{q}_{c1}  + \hat{q}_{c3} \Big). 
\end{align}
This now assumes that the variance of the feedforward operators:
\begin{align}
    V_2 = 3 \Delta^2 + 3({{1}\over{\eta \eta_{\text{g}}^2}}-1) + {{1}\over{\eta \eta_{\text{g}}^2}}({{1}\over{\eta_{\text{m}}}}-1),
\end{align}
is sufficiently small.

Finally we can consider the third mode. Its momentum operator directly after the CZ gate is given by:
\begin{align}
  \hat{p}_{3}' & = \sqrt{\eta} \sqrt{\eta_{\text{g}}^2} \Big(\hat{p}_{c3}  + \delta \hat{p}_{3} +  \hat{q}_{c2} +  \delta \hat{q}_{2} \Big) \nonumber \\
 & + \sqrt{1-\eta_{\text{g}}^2 \eta}\;(\delta \hat{q}_{v2}+\delta \hat{p}_{v3}). 
\end{align}
After linear amplification with $g = {{1}\over{\sqrt{\eta \eta_{\text{g}}^2 \eta_{\text{d}}}}}$, passing through the displacement loss and subsequent displacement by $\hat{p}_{c1o}$ and $\hat{p}_{c2o}$, the output momentum operator of the third mode is given by:
\begin{align}
  \hat{p}_{3}'' & = -\hat{p}_{c1} + \hat{p}_{c3}  + \delta \hat{p}_{3} + \delta \hat{q}_{2}  \nonumber \\
 & + \sqrt{{{1}\over{\eta_{\text{g}}^2 \eta}}-1}\;(\delta \hat{q}_{v2}+\delta \hat{p}_{v3}) \nonumber \\
 & - \sqrt{\eta_{\text{d}}}\sqrt{{{1}\over{\eta_{\text{g}}^2 \eta}\eta_{\text{d}}}-1}\; \delta \hat{p}_{v4} 
  + \sqrt{1-\eta_{\text{d}}}\; \delta \hat{p}_{\text{d}}. 
\end{align}
Similarly we can write the final position operator for the third mode as: 
\begin{align}
  \hat{q}_{3}'' & = -\hat{q}_{c1} - \hat{p}_{c2}  + \delta \hat{q}_{3}  \nonumber \\
 & + \sqrt{{{1}\over{\eta_{\text{g}}^2 \eta}}-1}\;\delta \hat{q}_{v3}
 \nonumber \\
 & - \sqrt{\eta_{\text{d}}}\sqrt{{{1}\over{\eta_{\text{g}}^2 \eta}\eta_{\text{d}}}-1}\; \delta \hat{q}_{v4} 
  + \sqrt{1-\eta_{\text{d}}}\; \delta \hat{q}_{\text{d}}, 
\end{align}
and hence we can write:
\begin{eqnarray}
\hat a_{3o} & = & {{1}\over{2}}(\hat{q}_{3}'' + i  \hat{p}_{3}'').
\end{eqnarray}
Although additional noise is added compared with Eq. \ref{3oc} the error correction feature of the circuit can still be seen from the fact that none of the noise terms in the output depend on the input. Again, even though there is still noise on the output, the noise does not ``build up" as we iterate but rather is ``refreshed" from new resource states at each round. Of course the extra noise introduced will need to be small and may require stronger squeezing of the source states.

\subsection{Logical Errors}

So far we have assumed that the measurement and feedforward always corrects to the right ideal solution, for example as in Eq.\ref{FF}.  However, as we saw from Eq.\ref{error}, even if the peaks are well-localized there is a non-zero probability, $1-P_j$, that they will fall outside their nominal domain.  If this happens, the wrong displacement will be fed forward.  For example, for GKP states if we wrongly identify a $q$ measurement falling on the right side of a peak with the neighbouring domain the ideal ``spikes" will be shifted by $\sqrt{2 \pi}$.  Similarly,  if we wrongly identify a $q$ measurement falling on the left side of a peak with the neighbouring domain the ideal ``spikes" will be shifted by $-\sqrt{2 \pi}$. After feedforward this will lead to bit-flip and/or phase flip errors in the output state. 


Given this, a better approximation for the feedforward operators is:
\begin{eqnarray}
\hat p_{c1o} &= &  \hat p_{c1} + \hat q_{c2} + \hat p_{e1} , \nonumber \\
\hat p_{c2o} &=& \hat p_{c2} + \hat q_{c1} +  \hat q_{c3} + \hat p_{e2},
\end{eqnarray}
where the $\hat p_{ei}$ are error operators with discrete outcomes $n \sqrt{2 \pi}$. If $n=0$ there is no error. This occurs with probability $P_i^{(0)} =  Erf[{{D}\over{2 \sqrt{2 V_i}}}]$. Errors happen when $n \ne 0$ and these occur with probabilities $P_i^{(n)} =  
(Erf[{{(|n|+1)D}\over{2 \sqrt{2 V_i}}}]-Erf[{{|n| D}\over{2 \sqrt{2 V_i}}}])$. The variances $V_i$ are the total noises on the detected quadratures, for example in the ideal case, from Eq.\ref{MO}, we have $V_1 = \langle \delta \hat q_{2}^2 \rangle +  \langle \delta \hat p_{1}^2 \rangle$ and $V_2 =  \langle \delta \hat q_{1}^2 \rangle +  \langle \delta \hat q_{3}^2 \rangle + \langle \delta \hat p_{2}^2 \rangle$.  The output operator can then be written more accurately as:
\begin{eqnarray}\label{3oe}
\hat a_{3o} & = & {{1}\over{2}}(\delta \hat q_3 + i \delta \hat p_3 + i \delta \hat q_2 + \hat p_{c2})\nonumber \\
& - & {{1}\over{2}}( \hat q_{c1} + \hat p_{e2} + i \hat p_{c1} +i \hat p_{e1}). 
\end{eqnarray}
Notice these mistakes do not change the noise properties, however now we can also track the signal errors as the system is iterated. A logical error encoding and correction scheme is required in order to correct the signal errors \cite{GLA06}. We believe that such multi-mode codes could also be tractable to analyse using our approach, however the complexity of the error tracking required may become more challenging.

\section{Discussion and Conclusion}

In this paper we have presented signal and noise analysis of both cat states and GKP states to highlight the generality of our approach. In particular the analysis of the preceeding section demonstrates the power and relative simplicity of this approach while also highlighting the necessity for the total noise variances to be small. Given we are grouping noise from several sources together and associating it with arbitrary logical states, it is important that we are in a regime where the noise affects different states in the same way. Fig.\ref{fig:GKP} shows that this is true for GKP states provided the squeezing is of order 10 dB or higher, as was observed previously. Whilst this may seem a bit restrictive, it should be remembered that for large quantum circuits to perform faithfully they inevitably need to operate in this high fidelity regime anyway \cite{MEN14,BAR19}. This also applies to quantum communication applications \cite{SCH24}. Hence, we expect our approach to prove useful in tracking the flow of noise, signal and errors in quantum computing circuits employing GKP and other Bosonic qubits. Finally, we have demonstrated the power these techniques can have to develop new schemes, with a loss tolerant generalisation of the simple teleportation circuit considered allowing for general loss rates for each component.

\section{Acknowledgement} 
TCR acknowledges useful discussions with Eli Bourassa, Ilan Tzitrin, Rafael Alexander and Joshua Guanzon. This work was partially supported by the Australian Research Council Centre of Excellence for Quantum Computation and Communication Technology (Project No.CE110001027). SNS was supported by the Sydney Quantum Academy, Sydney, NSW, Australia.

\end{document}